\documentclass[prl,twocolumn]{revtex4}
\usepackage{graphicx}

\def\eqa{\begin{eqnarray}}
\def\eea{\end{eqnarray}}
\newcommand{\eq}{\begin{equation}}
\newcommand{\ee}{\end{equation}}

\newcommand{\<}{\langle}
\renewcommand{\>}{\rangle}

\newcommand{\ua}{\uparrow}
\newcommand{\da}{\downarrow}

\newcommand{\al}{\alpha}

\newcommand{\cP}{ {\cal P} }

\begin{document}

\title{Unrestricted renormalized mean field theory of strongly correlated electron systems}
\author{Qiang-Hua Wang$^1$, Z. D. Wang$^{1,2}$, Yan Chen$^2$, and F. C. Zhang$^2$}
\affiliation{$^1$National Laboratory of Solid State
Microstructures, Nanjing University, Nanjing 210093, China}
\affiliation{$^2$Department of Physics \& Center of Theoretical
and Computational Physics, University of Hong Kong, Pokfulam Road,
Hong Kong, China}


\begin{abstract}
We generalized systematically the renormalized mean field theory
in the case of uniform states to the unrestricted case of general
inhomogeneous states with competing spin-, charge- and
superconducting orders. Applying the theory to high-$T_c$
superconductors, we discuss the issues of electronic
inhomogeneity, the superfluid density, and in particular the local
electron density of states. The results account for many
intriguing aspects of the phenomenology.
\end{abstract}

 \maketitle

As a prototype of strongly correlated electron systems, high-$T_c$
superconductors reveal rich phenomenologies. Apart from the
apparent competing anti-ferromagnetic and superconducting orders,
recent scanning tunnelling microscopy (STM) measurements have
suggested checkerboard charge-density-wave order in under-doped
NaCaCuOCl and BiSrCaCuO superconductors~\cite{checkerboard}. On
the other hand, earlier STM measurements already revealed
nano-scale spatial inhomogeneity of the quasi-particle energy
gap~\cite{gapdisorder}, but yet the local density of states (LDOS)
at very low energy are remarkably homogeneous, seemingly ignorant
of the gap inhomogeneity~\cite{robust}. The physics behind the
robust low energy quasi-particle states are intriguing. In short,
strongly correlated electrons may support complex phases and their
response to extrinsic impurities may be highly nontrivial. This
motivates us to develop a reliable effective theory that takes
proper care of the strong correlations and allows competing orders
and inhomogeneous electronic states from the starting point.

A widely adopted model for high-$T_c$ superconductors is the
$t$-$J$ model, \eqa H_{t-J}=-t\sum_{\langle ij\rangle \sigma
}(C_{i\sigma}^\dagger C_{j\sigma}+{\rm h.c.})+J\sum_{\langle
ij\rangle}S_i\cdot S_j, \eea where the constraint of no double
occupation at any site is implied. In order to fit the band
structure, hopping terms connecting next-nearest-neighbor sites
and further neighboring sites may be necessary but they do not
affect the following development of the theory. Recently the
$t-U-J$ model has also been used to study the Gossamer
superconductivity~\cite{tUJ}, in which the constraint of
no-double-occupation is relaxed but a Hubbard $U$-term is added,
\eqa H=H_{t-J}+U\sum_i n_{i\uparrow}n_{i\downarrow}.
\label{tuj}\eea The following development is formulated within
this general model, Eq.(\ref{tuj}), with obvious specific limits
to the $t-J$ and Hubbard models at our disposal.

A very powerful way to take proper care of the strong correlation
effects is to Gutzwiller project a trial wave-function,
$|\psi\rangle= P|\psi_0\rangle$, where $|\psi_0\rangle$ denotes a
free-particle many-body wave-function, with possible order
parameters, and $ P=\Pi_i (1-\alpha D_i)$ is a projection
operator. Here $D_i=n_{i\uparrow}n_{i\downarrow}$, and $\alpha$ is
a variational parameter between $0$ and $1$. The calculation can
be done by Monte Carlo methods~\cite{vmc}. However, it is limited
by the lattice size and the parameter space. Alternatively the
effect of Gutzwiller projection is taken into account
approximately in a re-normalized mean field theory (RMFT)~
\cite{rmft}, which agrees qualitatively with the Monte Carlo
result. The Gutzwiller projected $d$-wave BCS states appear to
capture the basic features of high-$T_c$ phase
diagram~\cite{vmc,rmft,plainvanilla}. In this paper, we develop a
systematic extension of the RMFT to the case of unrestricted
spin/charge densities. We then discuss the physical quantities
based on the unrestricted RMFT and explain the robust low energy
states observed in the STM.

Let $|\psi_0\>$ be a Hatree-Fock state which carries the bare
variational unrestricted order parameters. In the usual mean field
theory this wave function is used to calculate the expectation
value of the Hamiltonian, and the optimization with respect to the
order parameters leads to the mean field self-consistent
equations. For strongly-correlated electrons, however, this usual
type of approach misses the most important physics, namely, the
possible emergence of Mott insulating phases. This physics is
conveniently picked up by Gutzwiller projecting the Hatree-Fock
state, namely, the trial state is given by $|\psi\>= P|\psi_0\>$.
However, we are no longer justified to use a projector $P$ with
identical projection parameter $\al$ at each site, since the
effect of strong correlations depends on the electron density.
This motivates us to use a generalized projection operator
$\cP=\Pi_i \cP_i$ so that $|\psi\>=\cP |\psi_0\>$, with \eqa
\cP_i=y_i^{n_i}(1-\alpha_iD_i)=E_i+y_iQ_i+\eta_iy_i^2D_i,\eea
where $E_i=(1-n_{i\uparrow})(1-n_{i\downarrow})$ and
$Q_i=\sum_{\sigma}n_{i\sigma}(1-n_{i\bar{\sigma}})$ are the empty
and single occupation operators, respectively. These are standard
projection operators, satisfying $E_i^2=E_i$, $Q_i^2=Q_i$ and
$D_i^2=D_i$. Therefore \eqa
\cP_i^2=E_i+y_i^2Q_i+\eta_i^2y_i^4D_i.\eea Here $\eta_i=1-\al_i$,
and the fugacity $y_i$ is introduced so that the local charge
density is not changed by
projection~\cite{laughlin,anderson,gros}, or
$f_i=\<\psi_0|n_i|\psi_0\>=\< \psi|n_i|\psi\>/\<\psi|\psi\>$, the
advantage of which is that the variation of $\eta_i$ does not
influence the local charge density. (In the $t-J$ model
$\eta_i\equiv 0$.) We note that we are working in the
grand-canonical ensemble, which proves more convenient than the
usual canonical ensemble approach usually used in the case of
uniform state RMFT.

We shall use the shorthand notations $\< \cdot \>_0 \equiv
\<\psi_0|\cdot|\psi_0\>$ and $\<\cdot\> \equiv Z^{-1}
\<\psi_0|\cP\cdot \cP|\psi_0\>$ with the normalization factor $
Z=\<\psi_0|\cP^2|\psi_0\>=\< \cP^2\>_0$. Since $|\psi_0\rangle$
describes free particles, one can use Wick's theorem to obtain all
the contractions in $Z$. Under the Gutzwiller approximation,
inter-site correlations are ignored when expectation values of
projection operators are evaluated. One thus obtains $Z=\Pi_i
z_i$, with \eqa z_i=\langle
\cP_i^2\rangle_0=e_{0i}+y_i^2q_{0i}+\eta_i^2y_i^4d_{0i},\eea where
$e_{0i}=\langle E_i\rangle_0=1-f_i+d_{0i}$, $q_{0i}=\langle
Q_i\rangle_0= f_i-2d_{0i}$ and $d_{0i}=\langle
D_i\rangle_0=r_{i\uparrow} r_{i\downarrow}$. In the above
derivation, we have assumed the absence of on-site pairing and
spin flipping order parameters in the wave-function
$|\psi_0\rangle$. Similarly, we have $e_i=\langle
E_i\rangle=e_{0i}/z_i$, $q_i=\langle Q_i\rangle=q_{0i}y_i^2/z_i$,
and $d_i=\langle D_i\rangle=d_{i0}\eta_i^2y_i^4/z_i$. The fugacity
$y_i$ is determined by enforcing $f_i=q_i+2d_i$, and is a function
of $\eta_i$, $f_i$ and $d_{0i}$ (or equivalently $r_{i\sigma}$).
The calculation of $y_i$ is even unnecessary. Eliminating $y_i$ we
find $ d_ie_i/q_i^2=\eta_i^2d_{0i}e_{0i}/q_{0i}^2$, which
determines $d_i$ uniquely.

We now discuss the effect of projection on the local spin moment
directed in the $z$-direction. Since $S_i$ is already projective,
we have $\cP_iS_i\cP_i=y_i^2S_i$, and therefore \eqa m_i=\langle
S_i^z\rangle=\langle S_i^z\rangle_0 y_i^2/z_i=g_s(i)m_{0i},\eea
where $g_s(i)=q_i/q_{0i}$ is the re-normalization factor for the
variational spin moment $ m_{0i}=\langle
S_i^z\rangle_0=(r_{i\uparrow}-r_{i\downarrow})/2$. Similar
consideration applies to the spin-spin exchange, \eqa \langle
S_i\cdot S_j\rangle = g_s(i)g_s(j) \langle S_i\cdot
S_j\rangle_0.\eea

Using the identities \eqa
\cP_iC_{i\sigma}\cP_i=[y_i(1-n_{i\bar{\sigma}})+\eta_iy_i^3n_{i\bar{\sigma}}]C_{i\sigma},\\
\cP_iC_{i\sigma}^\dagger
\cP_i=[y_i(1-n_{i\bar{\sigma}})+\eta_iy_i^3n_{i\bar{\sigma}}]C_{i\sigma}^\dagger,\eea
we obtain \eqa \langle C_{i\sigma}^\dagger C_{j\sigma}\rangle =
g_{t\sigma}(i)g_{t\sigma}(j)\langle C_{i\sigma}^\dagger
C_{j\sigma}\rangle_0,\eea with the re-normalization factors \eqa
g_{t\sigma}(i)=(1-r_{i\bar{\sigma}})\sqrt{\frac{e_iq_i}{e_{0i}q_{0i}}}
+r_{i\bar{\sigma}}\sqrt{\frac{q_id_i}{q_{0i}d_{0i}}}.\label{gt}\eea
In Eq.(\ref{gt}) the first term arises from the hopping process
that does not encounter $\bar{\sigma}$-electrons at site $i$, and
the second is from that involving a double occupation. We note
that the re-normalization factors for spin-spin exchange and the
hopping reduce in disguise to the known results in uniform charge
density and uniform magnetically ordered states~\cite{rmft}. The
result in the non-magnetic case for the $t-J$ model was used for
granted in Ref.~\cite{liyq}.

The total internal energy $E=\<H\>$ can then be written as, \eqa
E=-t\sum_{\langle ij\rangle\sigma}
g_{t\sigma}(i)g_{t\sigma}(j)(\chi_{ij\sigma}^*+\chi_{ij\sigma})\nonumber\\
-\frac{3J}{8}\sum_{\langle
ij\rangle}g_s(i)g_s(j)(\chi_{ij}^*\chi_{ij}+\Delta_{ij}^*\Delta_{ij})\nonumber\\
+J\sum_{\langle ij\rangle}g_s(i)g_s(j)m_{0i}m_{0j}+\sum_i
Ud_i,\nonumber\eea where we have defined $\chi_{ij\sigma}=\langle
C_{i\sigma}^\dagger C_{j\sigma}\rangle_0$ and
$\chi_{ij}=\sum_\sigma \chi_{ij\sigma}$,
$\Delta_{ij}=\<B_{ij}\>_0$ with $B_{ij}=
C_{i\da}C_{j\ua}-C_{i\ua}C_{j\da}$, and used the standard
decomposition \eqa \langle S_i\cdot
S_j\rangle_0=m_{0i}m_{0j}-\frac{3}{8}(\chi_{ij}^*\chi_{ij}+\Delta_{ij}^*\Delta_{ij}).\eea
It is understood that $r_{i\sigma}$, $\chi_{ij\sigma}$ and
$\Delta_{ij}$ are determined by $\psi_0$, $d_i$ is determined by
$r_{i\sigma}$ and $\eta_i$, and therefore $E$ is eventually a
functional of $\psi_0$ and $\{\eta_i\}$.

The optimization of $E$ requires $\delta E/\delta\langle\psi_0|=0$
and $\partial E/\partial \eta_i=0$. The first condition leads to a
Sch\"{o}rdinger equation $H_{MF}|\psi_0\rangle = \lambda
|\psi_0\rangle$, with the mean field Hamiltonian \eqa
H_{MF}=\sum_{\< ij\>\sigma}\left( \frac{\partial E}{\partial
\chi_{ij\sigma}}C_{i\sigma}^\dagger C_{j\sigma}+{\rm
h.c.}\right)\nonumber\\
+\sum_{\langle ij\rangle}\left( \frac{\partial E}{\partial
\Delta_{ij}}B_{ij}+{\rm h.c.}\right)
+\sum_{i\sigma}\left(\frac{\partial E}{\partial
r_{i\sigma}}-\mu\right) n_{i\sigma},\eea where $\mu$ is the
chemical potential. This is a self-consistent unrestricted RMFT in
the sense that 1) $|\psi_0\rangle$ is determined by $H_{MF}$, and
so are the variational order parameters; 2) There is no
restriction on the spatial variation of the order parameters; 3)
The effect of strong correlation is reflected in the
re-normalization coefficients $g_{t\sigma}$ and $g_s$; 4) The
re-normalization coefficients depend on ${\eta_i}$, and the global
minimum of the energy $E$ is reached after the self-consistency of
the mean field theory and the optimization of ${\eta_i}$ are
simultaneously achieved. In the case of $t-J$ model one simply
sets $\eta_i=0$~\cite{ogata-extension}. The theory has the
advantage that the Mott physics is built in. In the
literature~\cite{gdft}, the unrestricted Gutzwiller projection has
been applied to the multi-band Hubbard models in dealing with
para- and ferromagnetic states in transition metals. Further
encoded in the present theory are the intriguing unrestricted
competing anti-ferromagnetic and superconducting orders, which are
most interesting in layered cuprate superconductors.

The mean field internal energy $\<H_{MF}+\mu N\>_0$ is different
from $E$, but a self-consistent $|\psi_0\>$ minimizes $E$ so that
$\delta E=\sum_n \epsilon_n \delta f_n+\sum_{n,n'}V_{n,n'}\delta
f_n \delta f_{n'}+\cdots $ where $\{\epsilon_n\}$ is the
single-particle spectrum of $H_{MF}$, $f_n$ is the Fermi-Dirac
occupancy of the $n$-th single-particle orbital, and $V_{n,n'}$ is
the residual two-body interaction kernel. The mean-field
single-particles are not necessarily the electrons themselves, but
are the bare Landau quasi-particles. The interpretation of these
states, and in particular, the excitation energy spectrum, would
be obscured if the charge density in $|\psi_0\>$ were different
from that in $|\psi\>$.

In what follows we shall apply the theory to the $t-J$ model,
relevant to the cuprates. First, we discuss the effect of the
projection on the observables. Since the theory is projective, the
order parameters in the RMFT are different from the measured ones.
The re-normalization in the spin moment $m_i=g_s(i)m_{0i}$ has
been derived previously. For the superconducting pairing order
parameters, it can be shown easily using the method described
above that $\tilde{\Delta}_{ij}=\<
B_{ij}\>=g_{\Delta}(i,j)\Delta_{ij}$, with
$g_{\Delta}(i,j)=\frac{1}{2}\sum_{\sigma}g_{t\sigma}(i)g_{t\bar{\sigma}}(j).$
In the charge-uniform nonmagnetic states, $g_\Delta =2x/(1+x)<1$
where $x$ (with $e_i=x$) is the hole-doping level away from
half-filling. This result is well-known in the literature, and is
argued to be related to the existence of pseudo-gap~
\cite{rmft,plainvanilla}. The pair-pair correlation function on
disconnected bonds is re-normalized in a similar fashion,
$\<B_{ij}^{\dagger}B_{kl}\>=g_{\Delta}^2\<B_{ij}^\dagger
B_{kl}\>_0$. This justifies the assignment of $\tilde{\Delta}$ as
representing the long-range off-diagonal order. However, the
pair-pair correlation on the same bond is re-normalized quite
differently, since $B_{ij}^\dagger B_{ij}$ projects out a singlet
spin pair on the bond $\<ij\>$,  $\< B_{ij}^\dagger B_{ij}\>
=-2\<S_i\cdot S_j-\frac{1}{4}n_i n_j\>=-2 g_s(i)g_s(j)\<S_i\cdot
S_j\>_0+\frac{1}{2}f_i f_j$. (In the $t-J$ model $Q_i=n_i$ so that
$q_i=f_i$.) Therefore we expect significant incoherent part in the
pair-pair susceptibility due to short-range spin correlations.

The superfluid density $\rho_s$ is related to the second order
response of free energy to a vector potential. At zero
temperature, it is related to the kinetic energy and is therefore
re-normalized in the same way as the hopping parameters are, so
that $\rho_s=g_{t\sigma}^2t\chi=2xt\chi/(1+x)$ in a uniform state,
where $\chi$ is the variational hopping order parameter in the
RMFT. (The expression is slightly changed in the presence of
longer-range hopping, but the re-normalization factor is the
same.) One realizes that this relation is consistent with the
result of elaborate gauge theory~\cite{gaugetheory}. Indeed
integrating over gauge fluctuations in the gauge theory
effectively restores, to some extent, the strong correlation
effects. The appealing feature of the projected density functional
theory is that no slave degrees of freedom, and therefore no gauge
fields are involved.

The factor $g_{t\sigma}(i)$ measures the overlap between the bare
quasi-particle state in the RMFT and a corresponding real electron
wave function at the given site. Therefore, the electron Green's
function may be written as $
G(i\sigma,j\sigma)=g_{t\sigma}(i)g_{t\sigma}(j)G_0(i\sigma,j\sigma)
+G_{inc}(i\sigma,j\sigma)$~\cite{anderson}, where the first
contribution is the coherent part from RMFT.  We suppressed the
energy arguments for brevity. Consider the on-site Green's
function. A corresponding decomposition applies for the spectral
function,
$A_\sigma(i)=g_{t\sigma}^2(i)A_{0\sigma}(i)+A_{inc,\sigma}(i)$.
Even though RMFT does not provide $A_{inc,\sigma}$ directly, some
properties of it can be argued, as discussed previously in the
literature~\cite{randeria}. We extend the discussion here for
generally inhomogeneous cases. Since $\int_{-\infty}^0 \sum_\sigma
A_\sigma(i)=\int_{-\infty}^0 \sum_\sigma A_{0\sigma}(i)=f_i$ but
$g_{t\sigma}(i)<1$, one concludes that there is incoherent
spectral weight in the {\em occupied} side. While the general
expression is quite complicated in general, it has a simple form
for a nonmagnetic site, which is given by $\int_{-\infty}^0
\sum_\sigma A_{inc,\sigma}(i)=(1-e_i)^2/(1+e_i)$. On the other
hand, the number of {\em unoccupied} states at site $i$ is given
by $\sum_\sigma \int d\omega \langle
C_{i\sigma}\delta(\omega-H)C_{i\sigma}^\dagger\rangle= 2e_i$,
where we dropped a contribution $q_i$ from the upper Hubbard band
, which is out of the Hilbert space of the $t-J$ model. To this
total unoccupied spectral weight, the contribution from the
coherent part of $G$, which does describe excitations below the
Mott gap, is given by \eqa \sum_\sigma
g_{t\sigma}(i)^2(1-r_{i\sigma})=\frac{2e_i}{1+4m_{0i}^2/(1-e_i^2)}.\nonumber\eea
Of particular interest is the nonmagnetic case ($m_{0i}=0$), in
which the above discussion predicts that the unoccupied spectral
weight is exhausted by the coherent part, in agreement with a
recent argument~\cite{randeria,note}. By continuity, one would
expect that the incoherent part in the occupied side vanishes
while approaching the Fermi level. The fact that incoherent
excitations are abundant in the occupied side preferably at higher
energies implies that 1) the nodal excitations at the Fermi energy
are quite robust against strong correlations and 2) the antinodal
excitations are largely incoherent due to strong correlations.
This seems to be consistent with the the energy-dependence of the
spectral function widely observed in angle-resolved photo-emission
measurements~\cite{arpesbackground}.

We now calculate the LDOS in the $t-J$ model with weak random
on-site potential, relevant to STM data. The inset of Fig.1(a)
shows the scalar impurity potential profiles. The main panel of
Fig.1(a) shows the coherent part of the LDOS calculated from the
RMFT at an average doping level $x=0.1$. The system carries
$d$-wave pairing order without spin moments. While the DOS at and
above the gap energy scale disperse from site to site, the low
energy DOS seems to be independent of position and ignorant of the
random impurities, in qualitative agreement with
experiments~\cite{gapdisorder}. In order to see whether this is a
universal feature of doped Mott insulators, we deliberately erased
the pairing order to find that the system evolves to a
staggered-flux (or $d$-density-wave) state, with circulating
currents whose chirality switches from plaquette to
plaquette~\cite{ddw}. The LDOS in this state is shown in Fig.1(b).
Now the low energy LDOS at different sites no longer merge
together, but those at an energy above the Fermi level does. Since
the DOS at the band-gap minimum vanishes linearly with energy in a
charge-uniform staggered-flux state in the same way as at the
Fermi level in a uniform $d$-wave paired state, we conclude that
the homogeneity in the low energy DOS in a $d$-wave superconductor
is merely a result of vanishing DOS in the parent uniform phase.
In other words, the strong correlation effect stabilizes the
$d$-wave pairing or density-wave states, and then some
quasi-particles are protected by the small phase space available
to elastic scattering.

A few remarks are in order. First, we are exaggerating the density
of impurities in Fig.1. The phase-space protection of the
quasi-particles is limited to weak disorder and low impurity
concentrations. At stronger disorder and denser impurities, the
inhomogeneities of the LDOS prevail at all energies. Second, in
comparison with the U(1) slave boson mean field
theory~\cite{u1mft}, our results are in better agreement with
experiments. This is because the U(1) mean field theory
underestimates the kinetic energy so that the holes are more
susceptible to impurity localization. In addition, as compared to
calculations without the projection and the self-consistency, the
LDOS in our case are more uniform against potential impurities in
space. The physics is best seen in the Mott insulator limit, where
the low-lying excitations are spinons, which are completely
ignorant to potential impurities. This means that strong
correlation effect reduces the charge susceptibility. Third, the
peak-to-peak gap in our case varies less severely than in
experiments~\cite{gapdisorder,robust}. We note that in order to
mimic the experimental data, gap-impurities were introduced
phenomenologically in a recent model~\cite{hirschfeld}. It would
be highly interesting to find a mechanism that could lead to such
gap impurities.

\begin{figure}
\includegraphics[width=8.5cm]{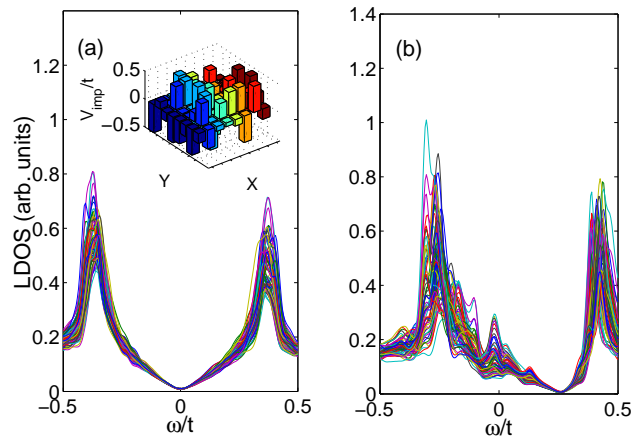}
\caption{ (Color online) The coherent part of the LDOS calculated
in a RMFT of the $t-J$ model with $t/J=3$ and at the average
doping level $e=10\%$. The curves as functions of energy at
different positions in an $8\times 8$ unit cell in a $400\times
400$ lattice are clumped together in order to see the difference
induced by the impurities at different energies. (a) LDOS in a
$d$-wave superconducting state; (b) LDOS in a staggered flux state
evolved from erasing the superconducting order. The inset in (a)
shows the impurity potential profiles.}
\end{figure}



\acknowledgements{We thank Dung-Hai Lee, P. W. Anderson, T. M.
Rice, Dieter Vollhardt, V. N. Muthukumar, and T. K. Ng for
stimulating communications. The work at Nanjing University was
supported by NSFC 10325416, 10429401 and 10021001, the Fok Ying
Tung Education Foundation No.91009, and the Ministry of Science
and Technology of China (973 project No: 2006CB601002), and the
work at the University of Hong Kong was supported by the RGC
grants of Hong Kong.}


\begin{references}
\bibitem{decomposition} The decomposition of the Green's function is general, but is different
from that proposed in Ref.\cite{anderson}.
\bibitem{note} This argument fails in one dimension where the
physics is governed by Luttinger-liquid behavior. The Gutzwiller
projection is expected to work better in higher dimensions.
\end{references}
\end{document}